\documentclass[prb,twocolumn,showpacs,preprintnumbers,amsmath,aps]{revtex4}
\usepackage{graphicx,hyperref}
\usepackage{bm}
\usepackage{subfigure}

\def\nbC{{\mathchoice {\setbox0=\hbox{$\displaystyle\rm C$}%
\hbox{\hbox to0pt{\kern0.4\wd0\vrule height0.9\ht0\hss}\box0}}
{\setbox0=\hbox{$\textstyle\rm C$}\hbox{\hbox
to0pt{\kern0.4\wd0\vrule height0.9\ht0\hss}\box0}}
{\setbox0=\hbox{$\scriptstyle\rm C$}\hbox{\hbox
to0pt{\kern0.4\wd0\vrule height0.9\ht0\hss}\box0}}
{\setbox0=\hbox{$\scriptscriptstyle\rm C$}\hbox{\hbox
to0pt{\kern0.4\wd0\vrule height0.9\ht0\hss}\box0}}}}
\def\nbQ{{\mathchoice {\setbox0=\hbox{$\displaystyle\rm
Q$}\hbox{\raise
0.15\ht0\hbox to0pt{\kern0.4\wd0\vrule height0.8\ht0\hss}\box0}}
{\setbox0=\hbox{$\textstyle\rm Q$}\hbox{\raise
0.15\ht0\hbox to0pt{\kern0.4\wd0\vrule height0.8\ht0\hss}\box0}}
{\setbox0=\hbox{$\scriptstyle\rm Q$}\hbox{\raise
0.15\ht0\hbox to0pt{\kern0.4\wd0\vrule height0.7\ht0\hss}\box0}}
{\setbox0=\hbox{$\scriptscriptstyle\rm Q$}\hbox{\raise
0.15\ht0\hbox to0pt{\kern0.4\wd0\vrule height0.7\ht0\hss}\box0}}}}
\def\nbT{{\mathchoice {\setbox0=\hbox{$\displaystyle\rm
T$}\hbox{\hbox to0pt{\kern0.3\wd0\vrule height0.9\ht0\hss}\box0}}
{\setbox0=\hbox{$\textstyle\rm T$}\hbox{\hbox
to0pt{\kern0.3\wd0\vrule height0.9\ht0\hss}\box0}}
{\setbox0=\hbox{$\scriptstyle\rm T$}\hbox{\hbox
to0pt{\kern0.3\wd0\vrule height0.9\ht0\hss}\box0}}
{\setbox0=\hbox{$\scriptscriptstyle\rm T$}\hbox{\hbox
to0pt{\kern0.3\wd0\vrule height0.9\ht0\hss}\box0}}}}
\def\nbS{{\mathchoice
{\setbox0=\hbox{$\displaystyle     \rm S$}\hbox{\raise0.5\ht0%
\hbox to0pt{\kern0.35\wd0\vrule height0.45\ht0\hss}\hbox
to0pt{\kern0.55\wd0\vrule height0.5\ht0\hss}\box0}}
{\setbox0=\hbox{$\textstyle        \rm S$}\hbox{\raise0.5\ht0%
\hbox to0pt{\kern0.35\wd0\vrule height0.45\ht0\hss}\hbox
to0pt{\kern0.55\wd0\vrule height0.5\ht0\hss}\box0}}
{\setbox0=\hbox{$\scriptstyle      \rm S$}\hbox{\raise0.5\ht0%
\hboxto0pt{\kern0.35\wd0\vrule height0.45\ht0\hss}\raise0.05\ht0%
\hbox to0pt{\kern0.5\wd0\vrule height0.45\ht0\hss}\box0}}
{\setbox0=\hbox{$\scriptscriptstyle\rm S$}\hbox{\raise0.5\ht0%
\hboxto0pt{\kern0.4\wd0\vrule height0.45\ht0\hss}\raise0.05\ht0%
\hbox to0pt{\kern0.55\wd0\vrule height0.45\ht0\hss}\box0}}}}
\def\nbZ{{\mathchoice {\hbox{$\sf\textstyle Z\kern-0.4em Z$}}
{\hbox{$\sf\textstyle Z\kern-0.4em Z$}}
{\hbox{$\sf\scriptstyle Z\kern-0.3em Z$}}
{\hbox{$\sf\scriptscriptstyle Z\kern-0.2em Z$}}}}

\begin{document}

\title{Avalanches and perturbation theory in the random-field Ising model}

\author{Gilles Tarjus$^a$}

\author{Matthieu Tissier$^{a,b}$} 
\affiliation{$^a$ LPTMC, CNRS-UMR 7600, Universit\'e Pierre et Marie Curie,
bo\^ite 121, 4 Pl. Jussieu, 75252 Paris c\'edex 05, France\\
$^b$ Instituto de F\'isica, Facultad de Ingenier\'ia, Universidad de la Rep\'ublica,
J. H. y Reissig 565, 11000 Montevideo, Uruguay.}

\date{\today}

\begin{abstract}
Perturbation theory for the random-field Ising model (RFIM) has the infamous attribute that it predicts at all orders a dimensional-reduction property for the critical behavior that turns out to be wrong in low dimension. Guided by our previous work based on the nonperturbative functional renormalization group (NP-FRG), we show that one can still make some use of the perturbation theory for a finite range of dimension below the upper critical dimension, $d=6$. The new twist is to account for the influence of large-scale zero-temperature events known as avalanches. These avalanches induce nonanalyticities in the field dependence of the correlation functions and renormalized vertices, and we compute in a loop expansion the eigenvalue associated with the corresponding anomalous operator. The outcome confirms the NP-FRG prediction that the dimensional-reduction fixed point correctly describes the dominant critical scaling of the RFIM above some dimension close to $5$ but not below.  

\end{abstract}


\maketitle

\section{introduction}

The random-field Ising model (RFIM)  is a paradigmatic model for describing the effect of quenched disorded on the critical behavior of systems belonging to a variety of fields,\cite{imry-ma75,nattermann98} from physics and physical chemistry to econophysics. It is also notorious for being one example where perturbation theory, \textit{i.e.}, an expansion around a Gaussian reference theory and the associated perturbative renormalization group, appears to fail completely. Indeed, perturbation theory predicts to all orders that the critical behavior of the RFIM in dimension $d$ is identical to that of the pure Ising model in dimension $d-2$, a property known as dimensional reduction.\cite{aharony76,grinstein76,young77,parisi79} However, the result is wrong in low dimensions, as rigorously shown for $d=3$.\cite{imbrie84,bricmont87}

We found an explanation for this dimensional reduction breakdown by means of the nonperturbative functional renormalization group (NP-FRG).\cite{tarjus04,tissier06,tissier11} Within the NP-FRG, the breakdown of dimensional reduction, and the associated spontaneous breaking of
an underlying supersymmetry,\cite{parisi79} are attributed to the appearance of a strong enough nonanalytic dependence, a ``cusp'', in the dimensionless renormalized cumulants of the random field at the fixed point. The NP-FRG predicts that such a cusp is present at the zero-temperature fixed point that controls the critical behavior of the model\cite{nattermann98} when the spatial dimension is lower than a nontrivial critical value, $d_{\text{DR}}\simeq 5.1$, whereas only weaker non-analyticities appear when $d>d_{\text{DR}}$. In consequence, dimensional reduction is broken below $d_{\text{DR}}$ but is valid above.\cite{tarjus04,tissier06,tissier11} (We recall that the upper critical dimension of the RFIM is $d=6$.)

Inspired by the related problem of an interface in a random environment,\cite{narayan92,BBM96,static_middleton,static-distrib_ledoussal} we also related this cuspy dependence of the cumulants and correlation functions to the presence of ``avalanches'', which are collective phenomena present in disordered systems at zero temperature.\cite{tarjus13} In equilibrium, such ``static'' avalanches describe the discontinuous change in the ground state of the system at values of the external source that are sample-dependent. Avalanches take place on all scales at the critical point, but whether or not they induce a cusp in the {\it dimensionless} renormalized cumulants of the random field at the fixed point depends on their scaling properties, and more specifically on the fractal dimension $d_f$ of the largest typical avalanches at criticality compared to the scaling dimension of the total magnetization.\cite{tarjus13} The difference between these two dimensions correspond to a new exponent $\lambda$ that determines whether cuspy perturbations are irrelevant ($\lambda>0$) around the cuspless fixed point (see also below).\cite{footnote_lambda} The computation of $\lambda$ via the NP-FRG shows that indeed $\lambda>0$ in the RFIM for $d>d_{\text{DR}}$.\cite{tarjus13,baczyk14,footnote_LR}

In this paper, we come back to the perturbation theory and show how to take into account the effect of the avalanches within its framework. Of course this is possible only so long as the cuspless fixed point exists, which basically means above $d_{\text{DR}}$. In a nutshell, what we achieve is setting up a perturbative scheme to compute the exponent $\lambda$. To do this one has to augment the conventional perturbation theory to include \textit{cuspy perturbations}. We illustrate the computation to the $2$-loop order and we find that in $d=6-\epsilon$, 
\begin{equation}
\label{eq_lambda_eps}
\lambda=1- \frac 12 \epsilon -\frac 5{36}\epsilon^2 +\mathcal O(\epsilon^3).
\end{equation}
This indicates that $\lambda$ decreases as $d$ decreases below the upper critical dimension of $6$. The above $2$-loop result goes to zero for $d\simeq 4.6$. This is a rough estimate, and we provide improved results below (see Fig.~\ref{fig_lambda}), but it already shows that the cuspless fixed point leading to the dimensional-reduction property for the dominant critical scaling behavior has a \textit{finite} range of validity below $d=6$. This finding therefore confirms, through a method that is more familiar to most theorists dealing with critical phenomena, the outcome of the NP-FRG approach.

The rest of the paper is organized as follows. In section II we introduce the model and set up the perturbation theory. In section III we proceed to the renormalization of the amplitude of the cuspy perturbation up to two loops. We discuss the results and their consequences for the cuspless fixed point in section III and we provide some concluding remarks in section IV.

\section{model and perturbation theory}

Our starting point is the field-theoretical description of the RFIM in terms of a scalar field $\varphi(x)$ in a $d$-dimensional space and an effective Hamiltonian, or bare action,
\begin{equation}
\begin{aligned}
\label{eq_ham_dis}
S[\varphi;h]= &\int_{x}\bigg\{\frac{1}{2}[\partial \varphi(x)]^2+ \frac{r_B}{2} \varphi(x)^2 + \frac{u_B}{4!} \varphi(x)^4 \bigg\} \\&
-\int_{x} h(x) \varphi(x) \,, 
\end{aligned}
\end{equation}
where $ \int_{x} \equiv \int d^d x$ and $h(x)$ is a random ``source'' (a random magnetic field) that is taken with a Gaussian distribution characterized by a zero mean and a variance $\overline{h(x)h(y)}= \Delta_B \delta^{(d)}( x-y)$.

\begin{figure}[htb]
  \centering
\includegraphics[width=.3\linewidth]{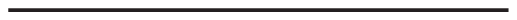}  \hfill \includegraphics[width=.3\linewidth]{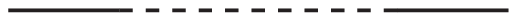}\hfill   \includegraphics[width=.3\linewidth]{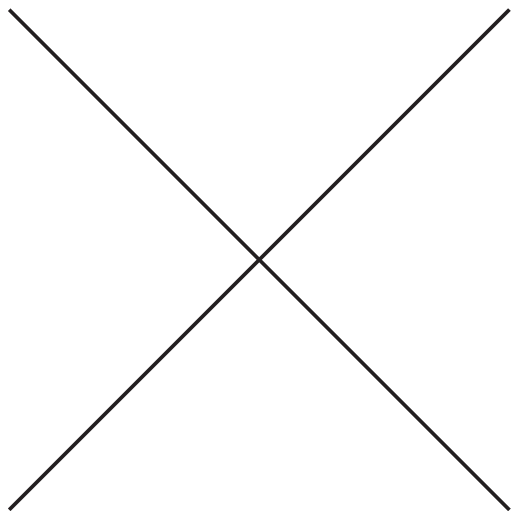} 
  \caption{Graphical representation of the free $1$-replica (connected) propagator, which is equal
    to $1/(q^2 +r_B)$, of the $2$-replica (also called ``disconnected'') propagator, $1/(q^2 +r_B)^2$, and of the bare $1$-replica vertex  $u_B$.}
  \label{fig-rules1}
\end{figure}

The equilibrium properties of the model are obtained from the average over disorder of the logarithm of the partition function,
which can be handled within the replica formalism by replacing the original problem by
one with $n$ replica fields, $\varphi_a(x)$, $a=1,2,\cdots,n$. After explicitly performing the average over the disorder in the partition function, one obtains a formulation in terms of a ``replicated action'',
\begin{equation}
\label{eq_replicated_bare}
\begin{aligned}
S_{\text{rep}}\left[\{\varphi_a\}\right ]&=  \int_{x}\bigg\{\sum_{a=1}^n\big[\frac 12 \partial \varphi_a(x) ^2+ \frac {r_B}2  \varphi_a(x) ^2+\\& \frac {u_B}{24}\varphi_a(x)^4\big]- \frac 12 \sum_{a,b=1}^n \varphi_a(x) \varphi_b(x)  \bigg\} \,,
\end{aligned}
\end{equation}
where we have set $\Delta_B\equiv 1$. For convenience, we will use the notation $\varphi_{B,a}$ in place of $\varphi_a$ to stress the bare nature of these fields as opposed to the renormalized ones. The replicated action contains a sum over one replica index and a sum over two replica indices. The 1-replica sum contains a quadratic part that gives rise to a free propagator and an interacting term $\propto u_B$ while the 2-replica sum is purely quadratic. The associated building blocks for a graphical representation of the perturbation expansion are shown in Fig.~\ref{fig-rules1}.

As well known, the perturbation theory of the above model leads to dimensional reduction where the zero-temperature fixed point in the presence of disorder reduces to the (finite-temperature) Wilson-Fisher fixed point of the pure model in dimension $d-2$.\cite{aharony76,grinstein76,young77}  This is a consequence of the existence  in the theory of a dangerously irrelevant operator which can be interpreted as the temperature. This induces a change in the canonical dimensions of the operators that is described by the $d\to d-2$ dimensional reduction. Our goal is then to determine the dimension of the ``cusp operator'' around this fixed point. As mentioned in the Introduction, this operator is the potential source of dimensional-reduction breakdown and it is physically associated with the presence of avalanches in the disordered system.

A linear cusp in the functional field dependence of the second cumulant of the renormalized random field corresponds, at the tree (mean-field) level, to  the following behavior of the $2$-replica, $2$-point proper vertex, $\Gamma^{(1,1)}_{B;ax,by}\equiv \delta^2 S_{\text{rep}}/[\delta \varphi_{B,a}(x)\delta \varphi_{B,b}(y)]$ with $a\neq b$, when $\varphi_{B,b} \to \varphi_{B,a}$:
\begin{equation}
\begin{aligned}
\label{eq_bare_cuspyvertex}
&\Gamma^{(1,1)}_{B;ax,by}-\Gamma^{(1,1)}_{B;ax,by}\vert_{\varphi_{B,b}= \varphi_{B,a}} \propto \\&
\delta^{(d)}(x-y)\big [ |\varphi_{B,a}(x) -\varphi_{B,b}(x)| +\mathcal O( [\varphi_{B,a} -\varphi_{B,b}]^2)\big ]\,,
\end{aligned}
\end{equation}
where $\Gamma^{(1,1)}_{B;ax,by}\vert_{\varphi_{B,b}= \varphi_{B,a}}=\Delta_B\equiv 1$ and the proportionality factor can still be a function of one of the fields, say $\varphi_{B,a}$. (Note that the above $2$-point proper vertex is symmetric under the permutation of $\varphi_{B,a}$ and $\varphi_{B,b}$ and should therefore be an even function of $[\varphi_{B,a}(x) -\varphi_{B,b}(x)]$.)
The above dependence is obtained by adding to the replicated action in Eq. (\ref{eq_replicated_bare}) the following ``anomalous'' contribution,
\begin{equation}
  \label{eq:anom-lag}
S_{\text{cusp}}=\frac {w_B}{4 }\int_x \sum_{a,b=1}^n \varphi_{B,a}(x) \varphi_{B,b}(x) |\varphi_{B,a}(x) -\varphi_{B,b}(x)|\,,
\end{equation}
where only the operator with the lowest canonical dimension, which corresponds here to a term cubic in the fields, is considered.

\begin{figure}[htb]
  \centering
    \includegraphics[width=.45\linewidth]{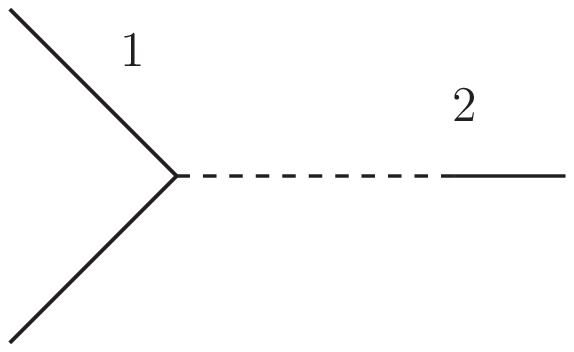}  \hfill \includegraphics[width=.45\linewidth]{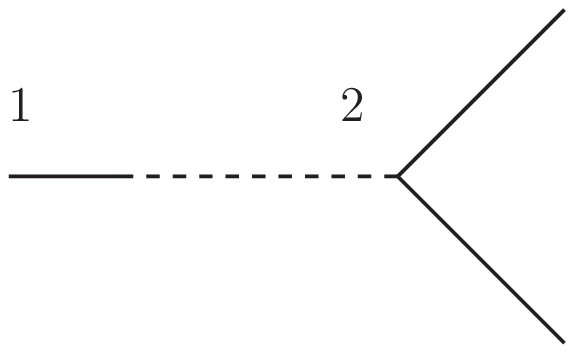}
  \caption{Graphical representation of the anomalous $2$-replica interaction vertices, which are equal to $w_B$ (left) and $-w_B$ (right) respectively.}
  \label{fig-rules2}
\end{figure}

The absolute value present in Eq. (\ref{eq:anom-lag}) implies that the sign of the associated $3$-point vertex depends on which field is larger. In what follows, in order to renormalize the vertex, we focus on two replicas (say, $a=1$ and $a=2$) and we consider the case where $\varphi_1>\varphi_2$. (Obviously, the final result is not modified if the other ordering is chosen.) As a consequence, there exists two anomalous $2$-replica interaction terms whose graphical representation is given in Fig.~\ref{fig-rules2}.

This completes the set of building blocks for the perturbation theory. Note finally that we concentrate on the massless theory and therefore discard the mass term $r_B$ from now on.

\section{Renormalization of the anomalous $3$-point vertex}

\begin{figure}[htb]
  \centering
  \includegraphics[width=.49  \linewidth]{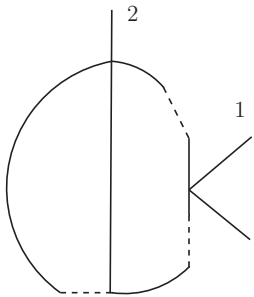} 
  \caption{Illustration of a diagram that can be discarded in the renormalization of $w$. Say we call $1$ the replica associated with the two external legs of the $1$-replica, $4$-point vertex on the right of the diagram and $2$ the replica associated with the external leg on the top of the diagram. Then, due to the connection through the $1$-replica connected propagator, all replicas entering in the anomalous vertex at the bottom of the diagram correspond to replica $2$. If we permute $1$ and $2$, the value of the diagram does not change sign. Therefore, it does not contribute to the renormalization of $w$.}
  \label{fig_notdiag}
\end{figure}

We now compute the beta function associated with the coupling constant $w$ of the anomalous term at two loops. To do so, we consider the $2$-replica, $3$-point vertex with two legs on replica $1$ and one on replica $2$. We keep only those diagrams that are linear in $w$ since we are interested in the eigenvalue $\lambda$ describing the flow in the cuspy direction in the vicinity of the (cuspless) fixed point. In addition, we can discard all diagrams where the two replicas entering in the anomalous vertex are actually identical because of the propagators joining them: see Fig.~\ref{fig_notdiag} for an explicit example. Indeed, such diagrams contribute with the same sign if we permute replicas $1$ and $2$, and therefore they do not renormalize $w$.

Finally, we only retain the terms having a maximum number of 2-replica (disconnected) propagators, as such diagrams dominate the infrared regime of the theory.\cite{aharony76,grinstein76,young77}

\subsection{One-loop calculation}

\begin{figure}[htb]
  \centering
 \includegraphics[width=.5\linewidth]{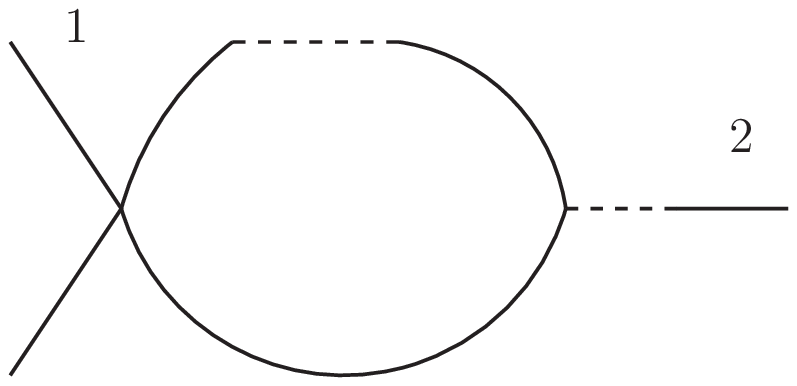}\hfill  \includegraphics[width=.4\linewidth]{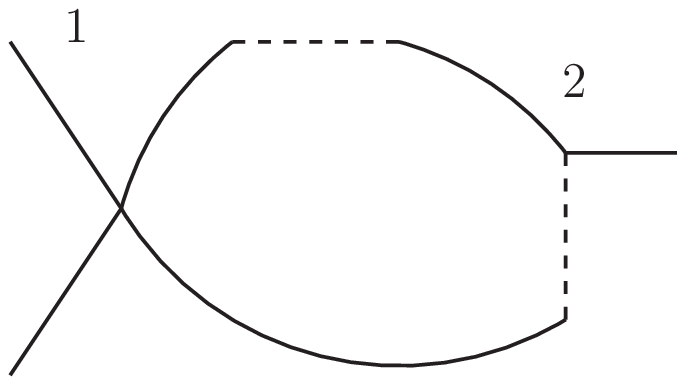}
  \caption{Feynman diagrams appearing in the one-loop calculation of the $2$-replica, $3$-point vertex for computing the eigenvalue $\lambda$.}
  \label{fig_one-loop}
\end{figure}

At one-loop order, only two diagrams contribute to the $2$-replica, $3$-point vertex for computing the eigenvalue $\lambda$. They are depicted in Fig.~\ref{fig_one-loop}.
It is easy to check that these two diagrams come with the same multiplicative factor. Moreover, when computing these two diagrams, one easily finds that the only difference results from the fact that the anomalous vertex appears with two legs with replica $1$ and one leg with replica $2$ in one diagram and two legs with replica $2$ and one leg with replica $1$ in the other diagram. Consequently, the two diagrams have equal magnitude and opposite sign. Their contribution therefore exactly cancel in the renormalization of $w$ and one needs to go to the two-loop level to get a nontrivial result.

\subsection{Two-loop calculation}

\begin{figure}[htb]
  \centering
    \includegraphics[width=.45  \linewidth]{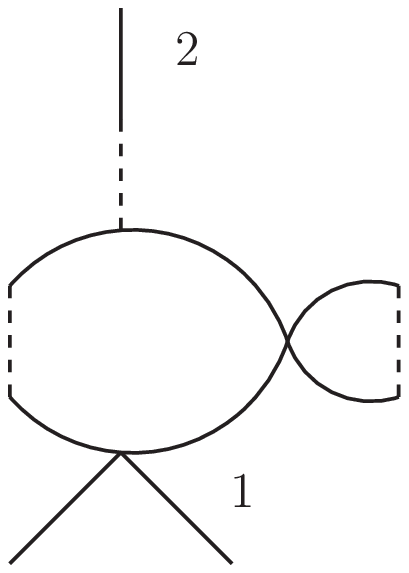} \hfill  \includegraphics[width=.45 \linewidth]{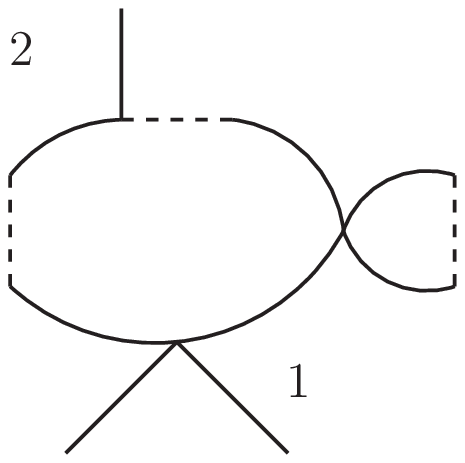}

 \includegraphics[width=.45  \linewidth]{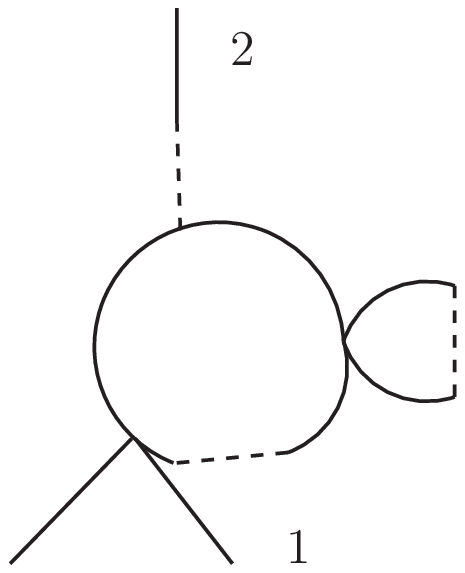} \hfill  \includegraphics[width=.45 \linewidth]{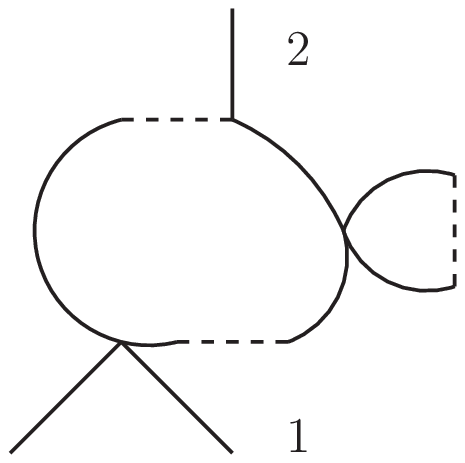}
  \includegraphics[width=.45  \linewidth]{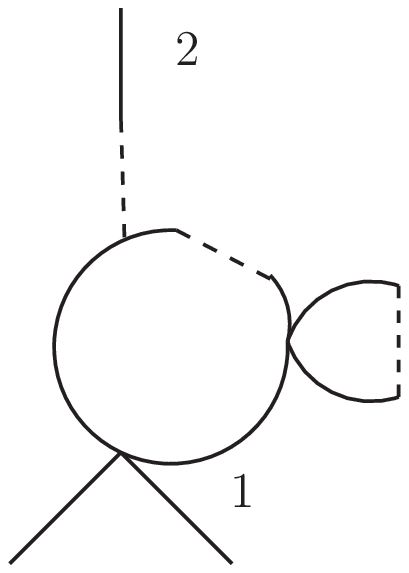} \hfill  \includegraphics[width=.45 \linewidth]{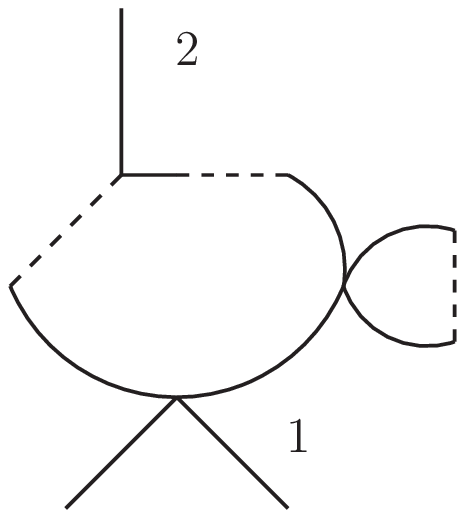}
  \caption{Two-loop diagrams of class $a$ contributing to the renormalization of $w$.}
  \label{fig_diaga}
\end{figure}

\begin{figure}[htb]
  \centering
  \includegraphics[width=.49  \linewidth]{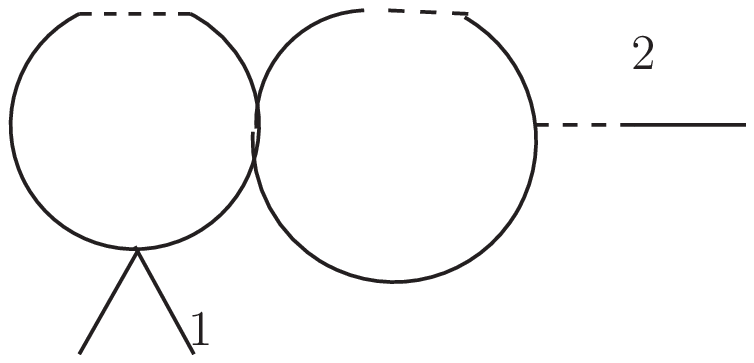} \hfill  \includegraphics[width=.49 \linewidth]{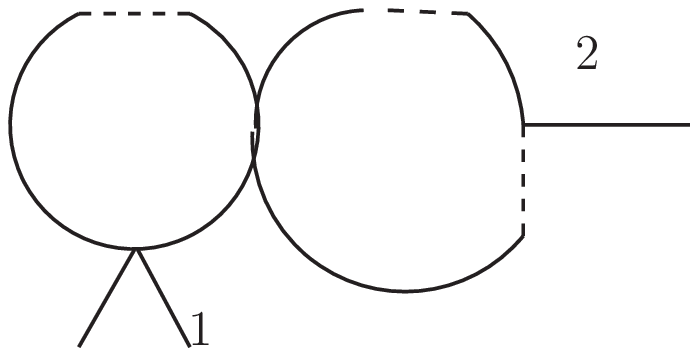}
  \caption{Two-loop diagrams of class $b$ contributing to the renormalization of $w$.}
  \label{fig_diagb}
\end{figure}

\begin{figure}[htb]
  \centering
  \includegraphics[width=.37  \linewidth]{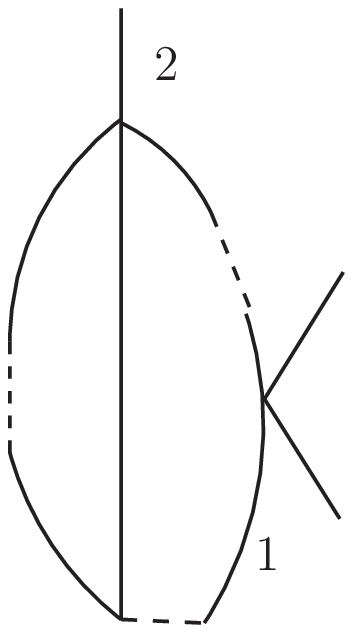} \hfill  \includegraphics[width=.37 \linewidth]{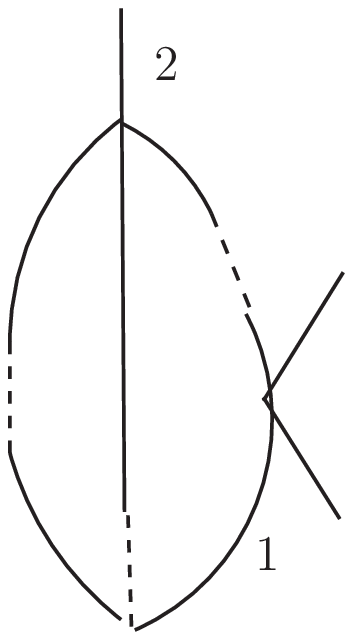}

  \includegraphics[width=.45  \linewidth]{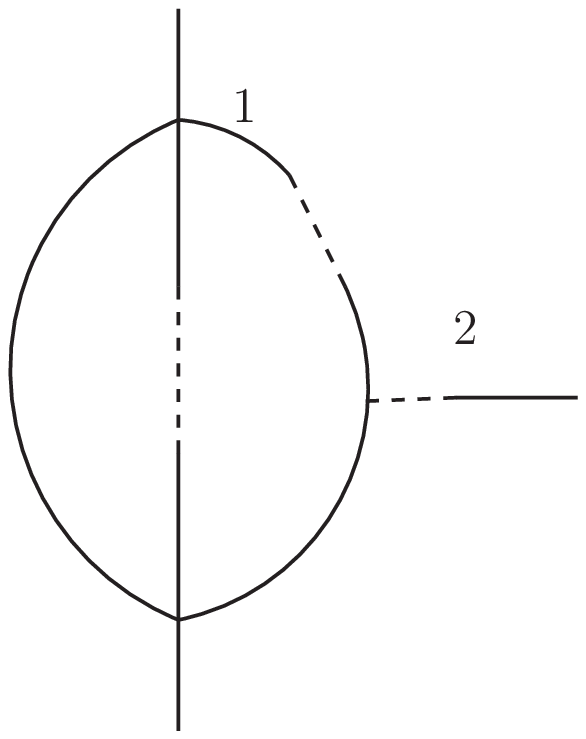} \hfill  \includegraphics[width=.4 \linewidth]{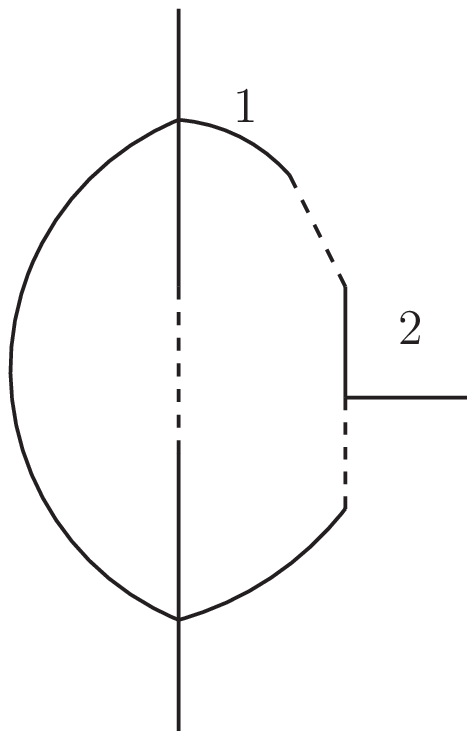}

  \caption{Two-loop diagrams of class $c$ contributing to the renormalization of $w$.}
  \label{fig_diagc}
\end{figure}

\begin{figure}[htb]
  \centering
 \includegraphics[width=.45  \linewidth]{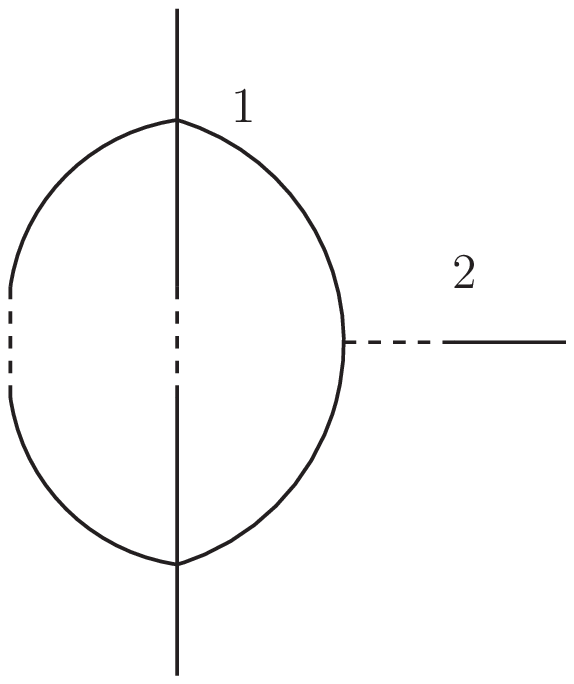} \hfill  \includegraphics[width=.35 \linewidth]{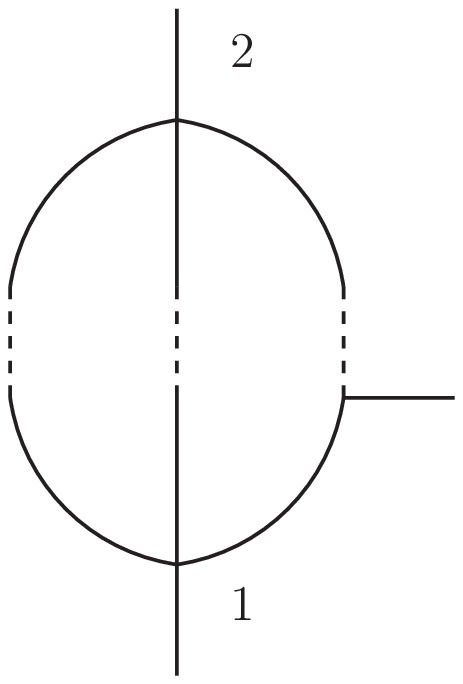}
  \caption{Two-loop diagrams of class $d$ contributing to the renormalization of $w$.}
  \label{fig_diagd}
\end{figure}

We list in Figs.~\ref{fig_diaga},  \ref{fig_diagb}, \ref{fig_diagc}, and \ref{fig_diagd} the diagrams that appear at two loops. 

It is interesting to note that almost all diagrams cancel by pairs, for the same reason as explained for the one-loop diagrams: this is the case for those of class $a$ to $c$ (Figs.~\ref{fig_diaga}-\ref{fig_diagc}). The only remaining contributions are those of the diagrams of class $d$ (Fig.~\ref{fig_diagd}). As a result, the relevant two-loop contribution to the $2$-replica, $3$-point vertex $\Gamma_{2,2\text{ loops}}^{(2,1)}$ for vanishing external momenta reads
\begin{equation}
  \label{eq_2loop}
  \begin{split}
&    \Gamma_{2,2\text{ loops}}^{(2,1)}=\\&-\frac 32 w_B u_B^2\int \frac{d^d
    p}{(2\pi)^d}\frac{d^d q}{(2\pi)^d}\frac 1{[p^2q^2(p+q)^2]^2}    
  \end{split}
\end{equation}
where $p$ is the external momentum of the leg on the replica 2.
The above integral is UV-divergent in $d=6$ and a standard calculation in dimensional regularization leads to the following leading term in an expansion for small $\epsilon=6-d$:
 \begin{equation}
  \label{eq_2loopdiv}
    \Gamma_{2,2\text{ loops},\text{ div}}^{(2,1)}=-3w_B u_B^2 \frac {K^2}\epsilon \,,
\end{equation}
where $K^{-1}=2^7 \pi^3$. Note that there is no contribution of order $1/\epsilon^2$, which is consistent with the fact that the one-loop contribution to the vertex is finite.

The above calculation was performed with bare coupling constants and bare fields. As usual, the divergences are absorbed by introducing renormalized quantities. Bare and renormalized quantities are related by renormalization factors in the following way:
\begin{align}
  u_B&=Z_u \mu^\epsilon\, u\\
w_B&=Z_w \mu^{-1+\epsilon/2}\, w\\
\varphi_{B,a}&=\sqrt {Z_\varphi}\, \varphi_a
\end{align}
where $\mu$ is a renormalization scale and where the powers of $\mu$ are chosen such that the renormalized coupling constants are dimensionless. 

The renormalization factors $Z_u$ and $Z_\varphi$ are given by the dimensional reduction. We only need here the first nonvanishing contribution for each of them, which, in the minimal subtraction scheme,\cite{zinnjustin89} reads
\begin{align}
  Z_u&=1+Z_u^{(1)}+\mathcal O(u^2),\\
  Z_\varphi&=1+Z_\varphi^{(2)}+\mathcal O(u^3),
\end{align}
with 
\begin{align}
Z_u^{(1)}&=  6   \frac {K u} \epsilon,\\
Z_\varphi^{(2)}&=-\frac 1 3 \frac {\left(   K u \right)^2}\epsilon \,.
\end{align}
This leads to the usual leading contributions for the beta function of the coupling constant $u$ and for the anomalous dimension, namely,
\begin{align}
  \beta_u=\mu \frac{\partial u}{\partial \mu}\Big |_{B}=-\epsilon
  u+6K u^2+\mathcal O(u^3)\,,\\
\eta=\mu \frac{\partial \log Z_\varphi}{\partial
  \mu}\Big |_{B}=\frac 23(K u)^2+\mathcal O(u^3) \,.
\end{align}

The renormalization factor $Z_w$ can now be derived by imposing that the renormalized vertex,
\begin{align}
  \Gamma^{(2,1)}_{r;2}=Z_\varphi^{3/2}(w_B+\Gamma_{2,2\text{ loops},\text{ div}}^{(2,1)}) \,,
\end{align}
is finite when expressed in terms of renormalized quantities. In the above expression the appearance of the field renormalization factor accounts for the fact that we consider a $3$-point renormalized vertex. A simple calculation then gives
\begin{equation}
  Z_w=1-\frac 52 \frac{\left(  K u \right)^2}\epsilon+\mathcal O(u^3)\, .
\end{equation}
We can thus obtain the beta function for $w$, expanded at linear order in $w$, as
\begin{align}
  \label{eq_beta_w}
  \beta_w&=\mu \frac{\partial w}{\partial \mu}\Big |_{B}\\
&=w\left(1-\frac\epsilon 2- \frac{\partial \log Z_w}{\partial u}\beta_u\right)\\
&=w\left(1-\frac\epsilon 2- 5 K^2 u^2+\mathcal O(u^3)\right)\,.
\end{align}
The eigenvalue $\lambda$ associated with the cuspy operator is then equal to $\lambda=1-\frac\epsilon 2- 5 (K u_\star)^2+\mathcal O(u_{\star}^3)$, where $u_\star$ is the value of the coupling constant at the fixed point. By using the fact that the Wilson-Fisher fixed point (obtained by dimensional reduction) is characterized by $K u_\star=\epsilon/6 + \mathcal O(\epsilon^2)$, one finally arrives at the two-loop expression given in Eq. (\ref{eq_lambda_eps}). Higher orders of the loop expansion could be obtained along the same lines.

\section{results and discussion}

\begin{figure}[htb]
  \centering
\includegraphics[width=\linewidth]{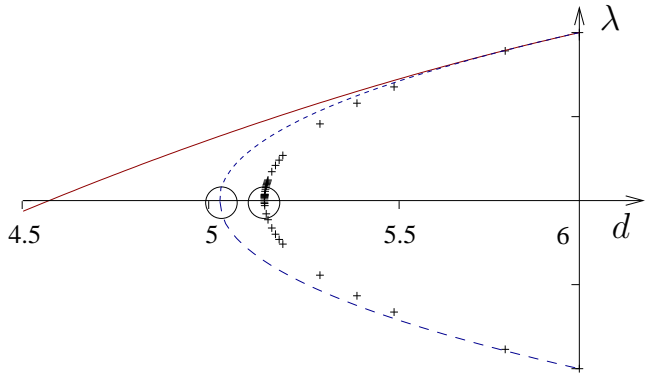}
  \caption{Variation with dimension $d$ of the eigenvalue $\lambda$ associated with the dimension of the cuspy operator at the cuspless fixed point. Full (red) line: two-loop result around the dimensional-reduction (Wilson-Fisher) fixed point. Dashed line: Approximate perturbative prediction for the merging of two cuspless fixed points. Crosses: NP-FRG predictions from Ref. [\onlinecite{tarjus13,baczyk14}]. The circles indicate the point where the two cuspless fixed points merge.}
  \label{fig_lambda}
\end{figure}

Taken at face value, the two-loop result for the eigenvalue $\lambda$ indicates that the latter decreases as the dimension $d$ decreases below the upper critical dimension $d=6$, where it is positive and equal to $1$, and goes through zero for $d\simeq 4.6$: see Fig.~\ref{fig_lambda}. The cuspless fixed point associated with dimensional reduction is therefore stable with respect to a nonanalytic cuspy perturbation only when $d>4.6$. In physical terms, this means that the avalanches that are the source of the cuspy behavior in some correlation functions and proper vertices are present but have a subdominant influence on the long-distance properties of the RFIM for a finite range of dimension below $6$, {\it i.e.}, so long as $\lambda>0$. Perturbation theory indicates that the cuspless dimensional-reduction fixed point is unstable below some dimension where the critical behavior of the RFIM should therefore be controlled by a different, presumably cuspy, fixed point.

The NP-FRG approach predicts that the cuspless dimensional-reduction fixed point of the RFIM actually disappears even before the eigenvalue $\lambda$ goes to zero: It does so for $d\simeq 5.1$ when $\lambda$ is very small but positive.\cite{tarjus13,baczyk14} The scenario involves an unstable conjugate cuspless fixed point that merges with the stable dimensional-reduction one in $d\simeq 5.1$ where a novel, cuspy, fixed point emerges. A related mechanism is also found for the $O(N>1)$ version of the random field model in a functional but perturbative renormalization-group analysis near the lower critical dimension $d=4$.\cite{baczyk14} Both in this case and in the NP-FRG approach, the conjugate unstable fixed point differs from the stable one by disorder-induced contributions that  involve multi-replica proper vertices and depend on field differences only. In the case of the RFIM described by Eq. (\ref{eq_ham_dis}), the eigenvalue $\lambda$ that is associated with the dimension of the cuspy operator, is given by $\lambda_{\text{unst}}=-(1-\epsilon/2) +\mathcal O(\epsilon^2)$ in $d=6-\epsilon$.\cite{tarjus13,baczyk14}

On may wonder what is the nature of the unstable cuspless fixed point in $d=6$ and whether one can set up a perturbation theory in  $d<6$. Guided by the NP-FRG results above mentioned, a natural candidate in the replica setting would be a fixed point differing from the Gaussian one by disorder-related vertices involving  $2$-, $3$-, etc., replicas and such that their functional dependence only includes differences between replica fields. Such terms correspond to adding a random potential to the disordered action, much like for the field-theoretical description of an elastic manifold pinned by a random environment.\cite{narayan92,BBM96,static_middleton,static-distrib_ledoussal} However, we have found no such fixed point that could be accessible by perturbative means in $d=6$ and the problem therefore remains open (for more details, see Appendix A).

It is nonetheless tempting to use the piece of information about the value of $\lambda$ for the NP-FRG unstable fixed point near $d=6$ (see above) and to combine it with the two-loop result derived in the previous section. One can for instance look for a polynomial form describing the merging as a function of $d$ of the eigenvalues $\lambda$ for two cuspless fixed point:
\begin{equation}
\epsilon(\lambda)= \epsilon_{DR}+\frac A2 (\lambda-\lambda_{DR})^2+\frac B6 (\lambda-\lambda_{DR})^3+\frac C{24} (\lambda-\lambda_{DR})^4 \,,
\end{equation}
where the (unknown) coefficients of the polynomial are determined by enforcing both Eq. (\ref{eq_lambda_eps}), when $\epsilon \to 0$ and $\lambda \to 1$, and the (putative) expression for $\lambda_{\text{unst}}$ to  a $\mathcal O(\epsilon^2)$, when $\epsilon \to 0$  and $\lambda \to -1$. The solution at this order is an even polynomial with $\lambda_{DR}=0$: $\epsilon(\lambda)=\frac{35}{36} -\frac{17}{18}\lambda^2-\frac 1{36}\lambda^4$. This leads to the curve plotted in Fig.~\ref{fig_lambda}. The two fixed points are then predicted to merge for $\epsilon_{\text{DR}}\simeq 0.97$, {\it i.e.}, for $d_{\text{DR}}\simeq 5.03$.  If instead of using $\lambda_{\text{unst}}$ to  a $\mathcal O(\epsilon^2)$ one just keeps $\lambda_{\text{unst}}=-1+\mathcal O(\epsilon)$, one obtains (with one less unknown coefficient) very close estimates: $\lambda_{DR}\simeq 0.029$ and $\epsilon_{\text{DR}}\simeq 0.94$, {\it i.e.}, $d_{\text{DR}}\simeq 5.06$.

We have thus shown that perturbation theory can still be useful for describing some aspects of the long-distance behavior of the RFIM. It remains valid for a finite interval of dimension near the upper critical dimension $6$. The new twist is that one should account for the effect of the avalanches, which are always present at zero temperature on all scales; this is made possible by studying operators that are nonanalytic in the field dependence. This extended perturbation theory, which leads to dimensional reduction but includes the effect of the avalanches on the critical behavior, definitely breaks down below a dimension that one can estimate to be close to $5$. This confirms the predictions of the nonperturbative functional renormalization group (NP-FRG). The latter of course has also the capability of describing the RFIM in lower dimensions, as shown in previous work.\cite{tarjus04,tissier06,tissier11,tarjus13bis} 

\appendix
\section{Searching for an unstable cuspless fixed point in $d=6$}

We briefly sketch here our attempt to find an unstable fixed point directly in $d=6$. 
A possible route is to look for a renormalized theory in $d=6$ with the following effective action
\begin{equation}
\begin{aligned}
\label{eq_unstable_dis}
\Gamma[\varphi;h,V]=\int_{x}\Big\{\frac{1}{2}[\partial \varphi(x)]^2- h(x) \varphi(x)  +\mathcal V_x(\varphi(x))\Big \}\,, 
\end{aligned}
\end{equation}
where $\int_x \equiv \int d^6 x$, the random field $h(x)$ is Gaussian with zero mean and a variance $\overline{h(x)h(y)}=\delta^{(d)}( x-y)$, and $\mathcal V_x(\varphi(x)$ is a random potential characterized by its cumulants. Alternatively, one can consider the cumulants of the renormalized random force $\mathcal F_x(\varphi(x))=-\delta \mathcal V_x(\varphi(x))/\delta \varphi(x)$. The random potential is associated with a statistical tilt symmetry which translates into the fact that the cumulants of the random force depends only on differences between fields. Note that a random effective action in terms of a local potential does not correspond to the most general spatial dependence but it is sufficient to illustrate our point.

In a replica setting, the above effective action leads to 
\begin{equation}
\label{eq_replicated_unstable}
\begin{aligned}
\Gamma_{\text{rep}}\left[\{\varphi_a\}\right ]&=  \int_{x}\Big\{\frac 1{2}\sum_a \partial \varphi_a(x) ^2-\frac {1}{2} \sum_{a_1,a_2} \varphi_{a_1}(x) \varphi_{a_2}(x) \\&
+\sum_{p=2}^{\infty}\frac {(-1)^{p-1}}{p!}\sum_{a_1,\cdots,a_p} D_p(\varphi_{a_1}(x),\cdots,\varphi_{a_p}(x))\Big\} \,,
\end{aligned}
\end{equation}
where the $D_p$'s are the cumulants of the random potential and, as already stated, depend only on field differences. 
The two first terms correspond to the Gaussian theory associated with the stable dimensional-reduction fixed point.

Through the exact RG equations for the above (running) effective action, we have checked that there {\it a priori} exists a consistent scheme with the cumulants of the random force $(-1)^{p-1}D_p^{(1,1,\cdots,1)}$ being exactly of power $p$ in the fields, {\it i.e.}, being linear combinations of Schur polynomials of degree $p$ of the $p$ variables $\varphi_{a_1},\cdots,\varphi_{a_p}$. (These combinations, we recall, should be invariant in any translation of the field, {\it i.e.}, $\varphi_{a}\to \varphi_{a}+\varphi_{0}$ $\forall a$.) 

To study this scheme in more detail it is convenient to use the transformation of variables proposed by Cardy in the context of the underlying supersymmetry of the RFIM:\cite{cardy85}
\begin{equation}
\begin{aligned}
&\varphi=\frac 12 \left [\varphi_1- \frac{\varphi_2+\cdots+\varphi_n}{n-1}\right ] \\&
\omega= \frac 1{2} \left [\varphi_1+ \frac{\varphi_2+\cdots+\varphi_n}{n-1} \right ]
\end{aligned}
\end{equation}
and 
\begin{equation}
\chi_\alpha=\sum_{a=2}^n c_{\alpha a} \varphi_a\,, 
\end{equation}
with  $\alpha=1,\cdots,n-2$ and $\sum_{a=2}^n c_{\alpha a}=0$.

The running effective action in Eq. (\ref{eq_replicated_unstable}) can then be rewritten, in the limit where the number of replicas $n \to 0$, as
\begin{equation}
\label{eq_cardy_unstable}
\begin{aligned}
&\Gamma_{\text{rep}}\left [\varphi,\omega,\{\chi_\alpha\}\right ]= \int_{x}\Big\{-\omega(x) \partial^2\varphi(x) -\frac {\omega(x)^2}{2} + \\&
\frac 12\sum_{\alpha=2}^{n-2} \partial \chi_\alpha(x)^2 + U(\{\chi_\alpha(x)\}) +\sum_{p=1}^\infty \omega(x)^pU_p(\{\chi_\alpha(x)\}) \Big \}\,.
\end{aligned}
\end{equation}
The potentials $U$ and $U_p$ are exactly zero when the random potential $\mathcal V$ is zero. One then recovers for the case $d=6$ Cardy's result,\cite{cardy85} {\it i.e.}, a theory with the peculiar feature of having $n-2$ free fields $\chi_\alpha$ when $n\to 0$. These fields can be interpreted as $2$ free fermionic ones, which leads back the supersymmetric formulation of Parisi and Sourlas\cite{parisi79} and the associated dimensional reduction by $2$.

We are however looking for an additional solution which, from the above result, should be a nontrivial, interacting theory at least in the sector of the $\chi_\alpha$'s. When using the symmetries and properties of the problem, it is easily derived that the sector of the $\chi_\alpha$'s can be studied independently of the two other fields $\omega$ and $\phi$ and that, in addition, the potential $U$ is a function of only $\sum_\alpha \chi_\alpha^2$. This amounts to studying a scalar field theory with an $n$-component field and $O(n)$ symmetry in the limit where $n=-2$. (We note in passing that the dimension of the cuspy operator in this interacting theory is $-(d-4)/2=-1$ whereas it is the opposite in the free theory.)

This $O(-2)$ model has a simple Gaussian-like critical behavior\cite{balian73,fisher73} but the existence of an associated fixed point has not been checked for arbitrary dimension. We have therefore looked for a nontrivial, interacting fixed point of the $O(-2)$ theory in $d=6$ through a nonperturbative RG calculation (within the so-called Local Potential Approximation)\cite{berges02} and found no physically acceptable one. Actually, we encountered the same situation everywhere between $d=4$ and $d=6$: the only acceptable fixed point of the $O(-2)$ seems to be the trivial Gaussian (free) one.

Whether or not the account of terms that become relevant below $d=6$, such as the $\varphi^4$ coupling constant $u$ would allow the putative unstable fixed point to become physical for $d<6$ is an open possibility that will however require a much more intensive investigation.

\end{document}